\documentclass[aps,prl,twocolumn,groupedaddress]{revtex4}

\usepackage{graphicx}

\begin{document}

%\preprint{}

\title{Ultra-stable lasers based on vibration insensitive cavities}


\author{J. Millo}
\author{D. V. Magalh\~{a}es}
\author{C. Mandache}
\author{Y. Le Coq}
\author{E. M. L. English}
\email[]{elizabeth.english@obspm.fr}
\author{P. G. Westergaard}
\author{J. Lodewyck}
\author{S. Bize}
\author{P. Lemonde}
\author{G. Santarelli}

%\thanks{}
\affiliation{LNE-SYRTE, Observatoire de Paris, CNRS, UPMC, 61
Avenue de l'Observatoire, 75014 Paris, France}


\date{\today}

\begin{abstract}

%We present the design of two different geometries of optical cavities (vertically and horizontally mounted).

We present two ultra-stable lasers based on two vibration
insensitive cavity designs, one with vertical optical axis
geometry, the other horizontal. Ultra-stable cavities are
constructed with fused silica mirror substrates, shown to decrease
the thermal noise limit, in order to improve the frequency
stability over previous designs.
%Analysis of the horizontal cavity
%includes full consideration of the effect of mirror tilt under
%horizontal acceleration.
Vibration sensitivity components measured are equal to or better
than 1.5 $\times$ 10$^{-11}$ per m\,s$^{-2}$ for each spatial
direction,  which shows significant improvement over previous
studies. We have tested the very low dependence on the position of
the cavity support points, in order to establish that our designs
eliminate the need for fine tuning to achieve extremely low
vibration sensitivity. Relative frequency measurements show that
at least one of the stabilized lasers has a stability better than
$ 5.6 \times 10^{-16}$ at 1 second, which is the best result
obtained for this length of cavity.

%Measurements of the two independent cavity stabilized lasers show
%a relative frequency stability of $\approx 7 \times 10^{-16}$
%between 1s and 10s.

\end{abstract}

\pacs{}
%\keywords{}

\maketitle

\section{1. Introduction}

Ultra-stable laser light is a key element for a variety of
applications ranging from optical frequency standards
\cite{ros,lud1}, tests of relativity \cite{muller}, generation of
low phase noise microwave signals \cite{bartels}, transfer of
optical stable frequencies by fiber networks \cite{newbury,jiang},
to gravitational wave detection \cite{virgo,ligo,lisa}. These
research topics, in particular cold atoms and single ion optical
frequency standards, have stimulated new approaches to the design
of Fabry-Perot reference cavities which are used to stabilize
lasers.
% computation of the Dick effect \cite{quessada}

For optical frequency standards with neutral atoms, the frequency
noise of state-of-the-art ultra-stable clock lasers sets a severe
limit to the clock frequency stability via the Dick effect
\cite{quessada}. Due to this limitation, the best reported Allan
deviations are more than one order of magnitude larger than the
ultimate quantum limit of these clocks \cite{lud1}. Improving the
laser frequency stability is therefore a prerequisite for
approaching this quantum limit.

\begin{figure}
\includegraphics[width=0.5\textwidth]{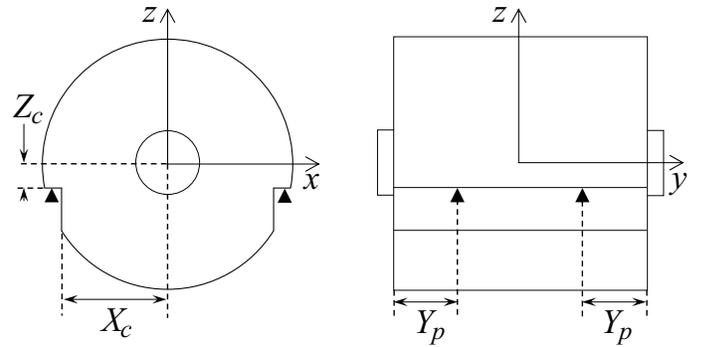}
\caption{ Front and side views of the horizontal cavity. The
optical axis lies along the $y$ axis. The four support points are
represented with black triangles. The positions of the cutouts for
support points are shown: $X_c$ with respect to the $yz$ plane,
and $Z_c$ with respect to the $xy$ plane. $Y_p$ is the distance
along the $y$ axis from the end of the cavity.\label{cavh}}
\end{figure}

One important issue for reducing the frequency noise of stabilized
laser cavities is to minimize the effects of residual vibration.
Vibration isolation systems can minimize the noise level, but
compact commercial systems are generally not sufficient to reach a
sub-Hz laser linewidth. One way to improve the spectral
performance of stabilized lasers is to reduce vibration
sensitivity by carefully designing the cavity geometry and its
mounting. Several groups have proposed and implemented low
vibration sensitivity cavities
\cite{bergquist,nazarova,ludlow2,web2}. A second important issue
is the reduction of thermal noise in cavity elements
\cite{numata}. The ultra-stable cavities presented here further
reduce both vibration sensitivity and thermal noise level, and
therefore improve cavity stability.

%In the classical approach, cylindrical optical cavities lie on
%V-blocks \cite{young}. In this configuration, the most important
%noise contribution comes from vibration-induced mechanical
%deformations. Indeed, a large sensitivity to vibration has been
%estimated by simulations and measurements to be in the region of
%3$\times$10$^{-10}$/(m.s$^{-2}$). The laser beam frequency is then
%compared to that of the reference cavity in vacuum and can be
%locked using the Pound-Drever-Hall technique \cite{pdh} onto a
%narrow resonance (typically 1-10 kHz wide).

\begin{figure}[t]
\includegraphics[width=0.5\textwidth]{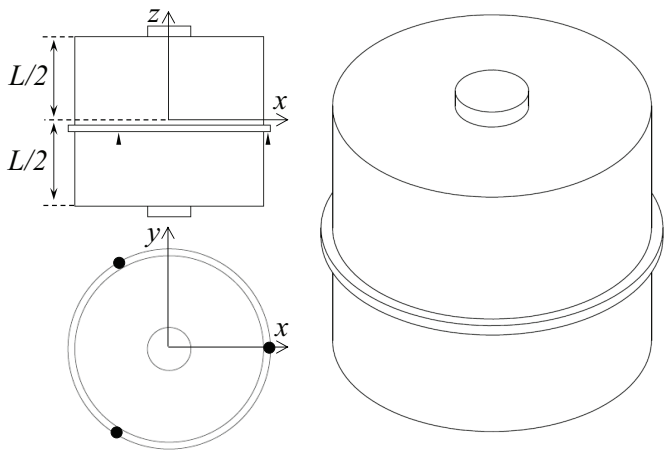}
\caption{\label{cavv} Top, side and isometric views of the
vertical cavity. The optical axis is aligned to the $z$ axis. The
three support-points are represented with black triangles or black
circles.}
\end{figure}

The two different optical cavities are designed based on the
results of extensive simulations using Finite Elements softwares.
The optical axis, which is also the axis of the spacer, is
horizontal for one cavity (Fig. \ref{cavh}) and vertical for the
other (Fig. \ref{cavv}). In each case, the position and size of
the cavity support points, and the effect of mirror tilt have been
analyzed. The constructed cavities have then been subjected to an
extensive study of the vibration response. Both cavity types
exhibit extremely low vibration sensitivity. Sensitivities are
equivalent to previous horizontal cavity designs
\cite{nazarova,web2} but with strongly reduced dependence on
support points position. The vertical cavity shows a much lower
sensitivity than previous vertical cavity designs \cite{ludlow2}.
Moreover, a significant improvement of the thermal noise level is
demonstrated here \cite{webster1,ludlow2} by using fused silica
mirror substrates, which minimize the contribution to thermal
noise due to the higher mechanical Q factor of this material in
comparison to Ultra Low Expansion glass (Corning ULE™).

%We develop two different designs of cylindrical cavities. The
%optical axis, which is also the axis of the spacer, is vertical
%for one design (Fig. \ref{cavv}) and horizontal for the other
%(Fig. \ref{cavh}). Through intensive use of a Finite Elements
%software package and multiprocessor computational power we have
%performed an extensive and refined study of the vibration response
%of these two cavities.

%Consequently, for a tight lock, the frequency noise spectrum of
%the laser is determined by the fluctuations of the optical length
%of the reference cavity.

%\begin{figure}
% \includegraphics[width=1]{fig_1.eps}\\
%  \caption{}\label{}
%\end{figure}

\section{2. Finite Elements Modelling of the Cavity}

\subsection{2.1.   General Considerations}

This analysis is restricted to the quasi-static response of
cavities as mechanical resonances are in the 10 kilohertz range,
while only low frequencies $< 100$ Hz are of interest in the
present experiment for application to optical atomic clocks.
Furthermore, with commercial compact isolation systems the
vibration level is still significant below $\sim$1 Hz where they
are not effective at reducing seismic noise.

In the finite element model, spacer and mirror substrates are
considered to be a single rigid body. The cavity geometry is
meshed with 125000 prism elements, where each prism has 6 nodes.
Finite element deformations are calculated within the elastic
limit. When the constrained cavity is accelerated, a length
variation is induced by elastic deformations. Careful design of
the cavity allows for compensation of this variation using
Poisson's effect \cite{pois} and cavity symmetries. Deformations
simulations have been done using the mechanical properties of ULE:
mass density (2210 kg\,m$^{-3}$), Young's modulus (67.6 GPa) and
Poisson's ratio (0.17).

%Vibration sensitivity of cavities is deduced by observing the
%deformations at the mirrors for a given acceleration value. We are
%not interested in observing the global deformations of the mirror,
%but in the displacement of the central region (where the laser
%beam is reflected). Therefore, two types of mirror deformations
%are important.

Vibration sensitivity of the cavities is deduced by observing the
deformation in the mirrors for a given acceleration value. The
displacement of the central region (where the laser beam is             %optical axis?
reflected) is of interest in the present study. Two types of
mirror deformation are important for both vertical and horizontal
cavities.

The first type is the mirror translation along the cavity axis, in
order to analyze the distance between mirror's centers. These
length variations exist if the cavity has no symmetry plane
orthogonal to the acceleration axis.

\begin{figure}
\includegraphics[width=0.4\textwidth]{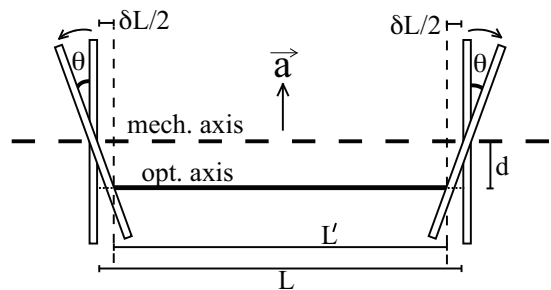}
\caption{ Schematic showing mirror tilt as a result of applied
transverse acceleration $\vec{a}$. This tilt occurs in both
horizontal and vertical cavities, and leads to changes in cavity
length $L$ as the optical axis is displaced from the mechanical
axis by a distance $d$. Note that the mirror tilt and $d$ are
exaggerated in this figure to illustrate this small effect
clearly. \label{tilta}}
\end{figure}

%For a real cavity, the mechanical and optical axes are misaligned
%when the cavity mirrors undergo a symmetric tilt of angle $\theta$
%(Fig. \ref{tilt}(a)). In the worst case for the cavities studied,
%these two axes are parallel and separated by a distance $d$, and
%the cavity length changes from $L$ to $L^{\prime}$. The special
%case (Fig. \ref{tilt}(b)) occurs when longitudinal vibration is
%applied to the horizontal cavity. Both mirrors are tilted by an
%angle $\theta$ in the longitudinal direction, with the result that
%optical and mechanical axes are no longer parallel.

The second type is the mirror tilt (Fig. \ref{tilta}), where the
mirrors are shifted through an angle $\theta$. For an ideal cavity
where optical and mechanical axes coincide, the tilt induced
length variation is a second order effect and can be neglected. In
a real cavity, mechanical and optical axes are not coincident due
to imperfections in the construction (e.g. mirror polishing,
spacer machining, contacting of the mirrors onto the spacer). A
worst case situation is considered in the present study, where the
optical and mechanical axes are parallel and displaced by a
distance $d$. The schematic in Fig. \ref{tilta} illustrates the
cavity deformation through mirror tilt and the optical length
change from L to L$^{\prime}$. Length variations become
proportional to both $d$ and the tilt angle. Consequently, tilt is
a first order effect on the cavity length and must therefore be
considered. Note that the mirrors' tilt angle $\theta$ is
extremely small in reality, so any change in the pointing
direction of the optical axis that may occur would be a second
order effect of $\theta$. Therefore these small movements in the
pointing direction of the optical axis do not change L
significantly.

%For an ideal cavity where optical and mechanical axes coincide,
%the tilt induced length variation is a second order effect and can
%be neglected. In a real cavity, mechanical and optical axes are
%misaligned due to imperfections in the construction (e.g. mirror
%polishing, spacer machining, contacting of the mirrors onto the
%spacer). For example, in the cavities studied these two axes are
%parallel and separated by a distance $d$.

%For a cavity with large vibration sensitivity coefficients of
%$10^{-10}$ an acceleration of 1 m.s$^{-2}$ (much greater than the
%acceleration that the cavity would undergo) produces a tilt angle
%of $\approx 10^{-9}$ rad. This corresponds to a shift of the
%optical axis of $10^{-10}$ m for a 100 mm long cavity. Compared
%with a typical beam diameter of a few hundred microns, this change
%in the pointing of the cavity due to mirror tilt does not
%significantly change, and therefore does not change the laser
%coupling into the cavity. Only the optical cavity length changes
%from $L$ to $L^{\prime}$ (Fig. \ref{tilta}), consequently tilt is
%a first order effect on the cavity length when the mechanical and
%optical axes are misaligned, and must therefore be considered.

For the horizontal cavity under longitudinal acceleration a
slightly different type of tilt configuration occurs, which will
be explained later on.

We can write the cavity length variations in the following way:
\begin{equation}\label{L}
\delta L/L = \vec{k}\cdot\vec{a}
\end{equation}
where $\vec{a}$ is the acceleration vector and $\vec{k}$ is the
vector of vibration sensitivities where the components are
expressed as:
\begin{equation}\label{ki}
\begin{array}{c}
  k_x = k_x^L + k_x^T(d) \\
  k_y = k_y^L + k_y^T(d) \\
  k_z = k_z^L + k_z^T(d) \\
\end{array}
\end{equation}
where $k_i^L$ and $k_i^T(d)$, are respectively the sensitivity
coefficients to the mirrors' translation and tilt.

\subsection{2.2.   Horizontal Cavity}

The horizontally mounted optical cavity configuration is 100 mm
long. For this cavity, the optical axis lies along the $y$ axis
and the spacer diameter is 100 mm (Fig. \ref{cavh}). The position
of the four support-points have been carefully calculated through
extensive simulations to design a cavity with very low vibration
sensitivity. The contact planes for these support points are
obtained by machining two square `cutouts' along the length ($y$
axis) of the cylindrical spacer. All four contact points are on
the same horizontal plane and are placed symmetrically around the
cavity as shown in Fig. \ref{cavh}.

Ideally, to be least sensitive to vibrations, the cavity spacer
should be a perfect cylinder supported at the horizontal mid plane
($xy$ axis), with contact points located on the surface of the
spacer. In this case the cavity and support points are completely
symmetric and so any acceleration will not induce mirror
translations.

%Practically, a cutout shoulder of at least 3 mm is required to
%support the cavity effectively. Therefore all simulations were
%calculated using $X_c = 47$ mm (\ref{cutout}). Due to the cutouts
%breaking the cavity symmetry, the vertical acceleration vector is
%no longer orthogonal to a symmetry plane. Hence, mirror
%translations contribute to the vertical vibration sensitivity. In
%this case and for some cutout geometries of the cavity, a specific
%position of support points ($Y_p$) suppress mirror translations,
%i.e. $k_z^L = 0$.

\begin{figure}
\includegraphics[width=0.5\textwidth]{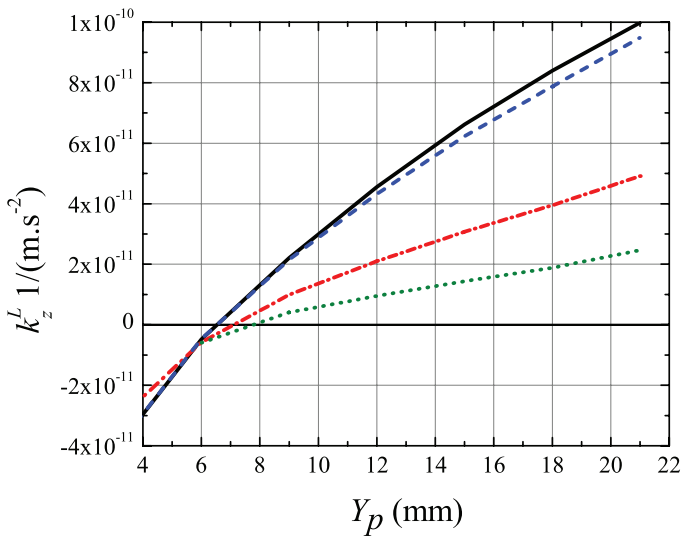}
\caption{ (Color online) Simulation showing the slope obtained for
$k_z^L$ for different cutout geometries. For each plot, $X_c = 47$
mm. Black solid line: $Z_c = 9$ mm, blue dashed line: $Z_c = 7$
mm, red dot-dashed line: $Z_c = 5$ mm, green dotted line: $Z_c =
3$ mm. \label{cutout}}
\end{figure}

Practically, we have assumed that a cutout shoulder of at least 3
mm is required to support the cavity effectively. Therefore most
simulations were calculated using $X_c = 47$ mm (Fig.
\ref{cutout}). Due to the cutouts breaking the cavity symmetry
about the $xy$ plane, cancellation of mirror translation due to
vertical acceleration is no longer guaranteed by the symmetry.
However, for some cutout geometries, a specific position of
support points ($Y_p$) can be found with the simulation which
suppresses mirror translations, \emph{i.e.} $k_z^L = 0$.

To make a practical adjustment of the support points' position
less critical, the slope of the acceleration sensitivity $k_z^L$
as function of the support points position $Y_p$ has to be
minimized. Simulations show that to achieve this, the cutouts must
be placed as close to the horizontal mid plane ($xy$ axis) as
possible.

\begin{figure}
\includegraphics[width=0.4\textwidth]{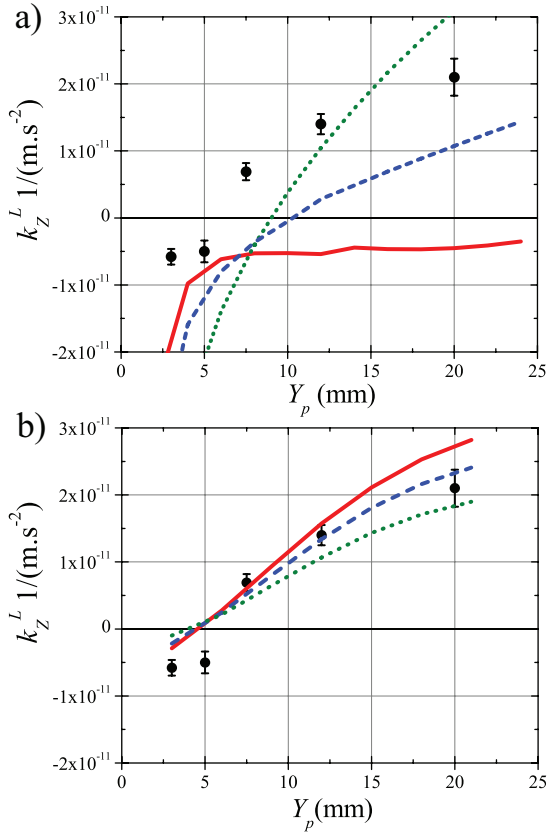}
\caption{ (Color online) Comparison between measurements and
simulation results for the vertical vibration sensitivity
(horizontal cavity) as a function of the support points distance
from mirror $Y_p$. Contact points are totally constrained (a) or
vertically constrained (b) with area 0.04 mm$^2$ (red line), 1
mm$^2$ (blue dashed line) and 4 mm$^2$ (green dotted line). The
support points of the cavity are 2 mm$^2$ (black
points).\label{vtcons}}
\end{figure}

Unfortunately, a vanishing sensitivity to vertical acceleration
$k_z^L = 0$ cannot be achieved for every cutout geometry.
Furthermore, simulations performed using two different models for
the support points have shown quite different results, as can be
clearly seen in Fig. \ref{vtcons}. We have assumed two mandatory
requirements for a good cavity design; firstly the existence of a
cancellation position $Y_p$ for all models of the support points,
and secondly, a good agreement between the position of the
cancellation position $Y_p$ for all models of the support points.
Based on these criteria we have excluded geometries with $Z_c <
3$mm as they do not show a cancellation position.

As an additional requirement, the optimum position of the support
points $Y_p$ where $k_z^L =0$ must also correspond to low values
for all other sensitivity coefficients. Concerning the sensitivity
to vertical acceleration, this means that we want to minimize tilt
of the mirrors so that $k_z^T$ is vanishingly small. This
requirement also means that we want $k_x$ and $k_y$ close to zero.
Since symmetries are insuring that $k_x^L$ and $k_y^L$ are both
zero, we are focused on cancelling $k_x^T$ and $k_y^T$. For each
coefficient, the aim was to achieve vibration sensitivity below
$10^{-11}$ (m\,s$^{-2}$)$^{-1}$ for a putative offset of the
optical axis $d = 1$ mm.

%When the spacer asymmetry due to cutouts is minimized, the
%longitudinal coefficient ($k_y$) also becomes smaller
%independently of the position of the support points. The optimum
%position of the support points where $k_z^L$ has a lowest value
%must also correspond to low values of all the other coefficients,
%which are due to mirror tilts ($k_z^T$, $k_x$, $k_y$ below $1
%\times 10^{-11}$ for $d = 1$).

%The coefficient due to mirror translation $k_z^L$ can vanish for a specific position of the support-points.

%The chosen geometry parameters $X_c = 47$ mm and $Z_c = 3$ mm take
%into account all these considerations; the 3 mm shoulder required
%to support the cavity, the existence of a zero-crossing for the
%two support points models, a low slope around the zero crossing as
%a function of the position of support points ($Y_p$) and minimized
%mirror tilts.

\begin{figure}
\includegraphics[width=0.5\textwidth]{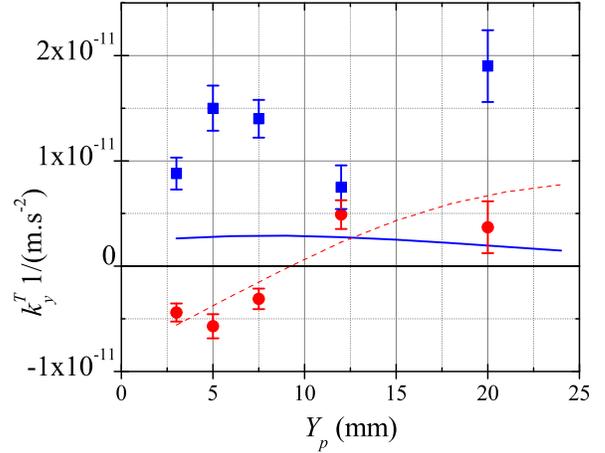} %trans_long.eps
\caption{(Color online) Horizontal vibration sensitivity of the
horizontal cavity as a function of $Y_p$ for $d = 1$ mm and
contact point size 2 mm$^2$. Transverse vibration sensitivity
measurements $k_x^T$ (red circles) and simulation (red dashed
line) and longitudinal vibration sensitivity measurements $k_y^T$
(blue squares) and simulation (blue line) are shown.
\label{horiz2}}
\end{figure}

With these considerations in mind, we have simulated a large
number of geometries and reached the conclusion that $X_c = 47$ mm
and $Z_c = 3$ mm was the best compromise. In the following we
present the results of the simulation obtained with this optimized
geometry, which are also shown in Figs. \ref{vtcons} and
\ref{horiz2}.

Figure \ref{vtcons} shows the sensitivity to vertical acceleration
$k_z^L$ as a function of the support points position. Simulations
have been performed for a range of contact point sizes for two
different models: totally constrained and only vertically
constrained. This figure illustrates that for this cavity design,
the coefficient $k_z^L$ is strongly dependent on the contact
model: the size of the contact points and the constraints placed
on the degrees of freedom.

When the contact points are constrained in all directions, a small
size gives a very low $k_z^L$ regardless of longitudinal position
$Y_p$ of contact points. For larger surface sizes, a solution with
$k_z^L = 0$ does exist for $Y_p \approx 10$ mm (Fig.
\ref{vtcons}). When the contact with the cavity is only vertically
constrained, the insensitive solution is independent of the
support-point size.

The vertical acceleration also induces tilt in the mirrors but the
sensitivity coefficient $k_z^T$ is independent of the contact
model. Assuming a rather large misalignment $d = 1$ mm, the
dependence on the support-point position is low
$\sim1.3\times10^{-12}$ (m\,s$^{-2}$)$^{-1}$ per mm, with $k_z^T =
0$ for $Y_p \sim 10$ mm.

Transverse ($k_x$) and longitudinal ($k_y$) components depend on
the position of the support-points, but are virtually independent
of the contact point size (Fig. \ref{horiz2}). Note again that due
to symmetry, only tilt of the mirrors contributes to both
horizontal sensitivities $k_x$ = $k_x^T(d)$ and $k_y$ =
$k_y^T(d)$. We find that the transverse component can be zeroed
for the support-point positions set at $Y_p \approx 9$ mm with a
slope $\approx 2 \times 10^{-12}$ (m\,s$^{-2}$)$^{-1}$ per mm.

%The acceleration applied in these measurements can cause unequal
%restoring forces on the cavity, accounting for some disagreement
%between simulation and measurements in Fig. \ref{horiz2}. However,
%the larger disagreement for the longitudinal measurements is still
%unexplained. It may be due to the fact that the acceleration is
%applied along the direction of the optic axis in the longitudinal
%case, whereas it is applied orthogonal to the optic axis in the
%other cases.

\begin{figure}
\includegraphics[width=0.4\textwidth]{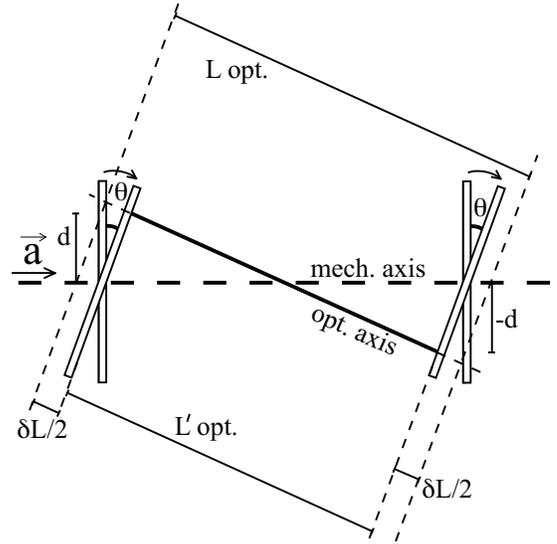}
\caption{ Schematic showing mirror tilt as a result of applied
longitudinal acceleration $\vec{a}$ occurring only in the
horizontal cavity. This leads to changes in optical length $L$, as
the mechanical axis and optical axis are misaligned and not
parallel. \label{tiltb}}
\end{figure}

When an axial vibration is applied ($k_y$), the two mirrors are
tilted in the same direction (Fig. \ref{tiltb}). The worst
misalignment to consider is when the mechanical and optical axis
are no longer parallel. In this case, we characterize this by a
displacement $d$ of the optical axis on one of the mirrors, and
$-d$ on the other. The longitudinal vibration sensitivity
component $k_y = k_y^T (d)$ also depends on $d$. In our chosen
geometry, we have found that this component can not be zero,
however for $d = 1$ mm it is always below $3 \times 10^{-12}$
(m\,s$^{-2}$)$^{-1}$ and therefore negligible (Fig. \ref{horiz2}).

%The choice of geometry parameters $X_c = 47$ mm and $Z_c = 3$ mm
%were also dictated by the consideration that regardless of the
%contact point model, it is always possible to find an optimum
%support spacing where $k_z$ has a low value of ($ \approx 5 \times
%10^{-12}$/(m.s$^{-2}$)). This optimum spacing also corresponds to
%a low value of $k_x$ ($<1 \times 10^{-11}$/(m.s$^{-2}$)).

\subsection{2.3.   Vertical Cavity}

The vertically mounted cavity geometry is 100 mm long, and the
optical axis lies along the $z$ axis (Fig. \ref{cavv}). The
contact plane is obtained by machining a central `shoulder' in the
spacer. The cavity is constrained by three equidistant support
points as shown in Fig. \ref{cavv}. This configuration allows the
distribution of equal restoring forces from the support to the
cavity.

The cavity has cylindrical symmetry around the optical axis ($z$
axis), therefore $k_x^L$ and $k_y^L$ vanish. Rigorously, the three
support-points break the rotational symmetry, however, simulation
results indicate that this has a negligible effect on vibration
sensitivity. The lack of an exact rotational symmetry means that
there is no geometry for which $k_x^T$ and $k_y^T$ can
simultaneously be zero. The magnitude of the horizontal vibration
sensitivity coefficients $k_x^T$ and $k_y^T$ depend on the
diameter:length ratio of the spacer.

\begin{figure}
\includegraphics[width=0.4\textwidth]{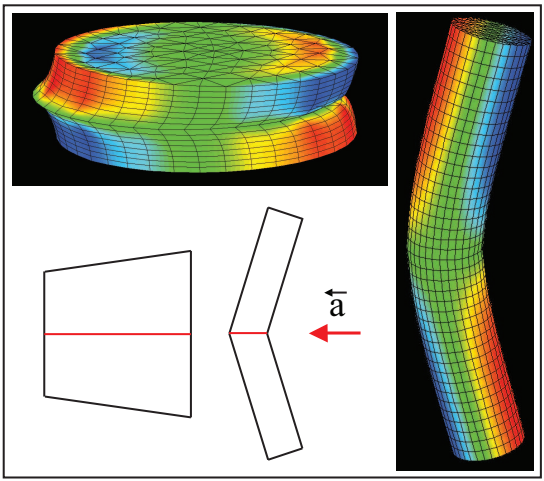}
\caption{(Color online) Deformation of a cylinder with ULE
properties when acceleration is applied to the center (arrow). For
the two different diameter:length ratios, the ends of each
cylinder are tilted in different directions (explained in text).
Color scale illustrates vertical displacement only, where blue is
negative and red is positive with respect to the center of the
non-deformed cylinder. Deformations are amplified by a factor of
$\approx 10^{6}$. \label{poisson}}
\end{figure}

When the diameter:length ratio is small, the cylinder deformation
is dominated by bending about the center. When it is large the
deformation induced by the Poisson effect dominates. In each case,
the ends of the cylinder are tilted in opposite directions and
therefore change the mirror tilt of the cavity (Fig.
\ref{poisson}). By choosing the correct ratio, it is possible to
cancel out the mirror tilt. Fixing the spacer diameter at 110 mm,
simulations indicate that a 100 mm long spacer is optimal,
minimizing both horizontal components $k_x^T$ and $k_y^T$ at $\sim
1\times10^{-12}$ (m\,s$^{-2}$)$^{-1}$ for $d = 1$ mm.

The vertical vibration sensitivity component depends on the
position of the contact plane and the geometry of the supporting
shoulder. This component is nearly vanishing when the contact
plane is optimally positioned, in this case 3 mm below the center
of the spacer for the chosen geometry of the shoulder as shown in
(Fig. \ref{cavv}). As a result, length variations of the upper
part of the spacer compensate exactly those of the lower part
\cite{Taylor}. Due to this geometry of the spacer and the forces
applied, the mirrors are translated without any mirror tilt.

\section{3.   Experimental Set-up}

Based on the results of these comprehensive simulations, two
ultrastable optical cavities have been designed and constructed;
one horizontal, the other vertical.

The spacers of the two cavity configurations are machined from ULE
glass rods. The wavelength range of the high reflection coating
mirrors allow operation at both 1064 nm and 1062.5 nm (Nd:YAG and
Yb doped fiber laser). Each cavity has been optically contacted
with a flat mirror and a concave mirror of radius of curvature 500
mm. Both cavities show a finesse of $\approx 800000$ and a fringe
contrast better than 50\%.

The substrates of the mirrors are made from fused silica to reduce
the contribution of thermal noise floor \cite{numata,notcutt}. For
the present geometry, this limit is estimated to have a flicker
noise floor of $\approx 4 \times 10^{-16}$ for a 100 mm long
cavity with fused silica mirrors, dominated by the thermal noise
of the high reflection coatings. The expected improvement compared
to an all ULE cavity is greater than a factor of 2
\cite{ludlow2,web2}.

%For an all ULE cavity this limitation is $\approx 1 \times
%10^{-15}$, dominated by the noise contribution of the mirror
%substrate \cite{ludlow2,web2}.

%\begin{figure}
%\includegraphics[width=0.5\textwidth]{fig6.eps}
%\caption{ Comparison between measurements (red circles) and
%results of simulations (blue diamonds) for the axial vibration
%sensitivity (horizontal cavity) as a function of $Y_p$ for $d = 1$
%mm.\label{fig6}}
%\end{figure}

However, fused silica shows a larger coefficient of thermal
expansion (CTE) than ULE. Consequently, the overall effective CTE
of the cavity is much larger than that of an all ULE cavity and
the zero thermal expansion coefficient is shifted to well below
$0^{\circ}$C, instead of $10-20^{\circ}$C for an all ULE cavity.
This increased temperature sensitivity requires a more
sophisticated design of the cavity environment. A high thermal
shielding factor coupled with a tight temperature control is
necessary to minimize the impact of environmental temperature
fluctuations.

%The thermo-mechanical setup of the vacuum cavity enclosure
%includes two independent vacuum chambers. This double shell avoids
%water condensation on windows when the internal chamber is cooled
%below the condensation point. Inside the inner chamber three
%thermal shields, polished and gold-coated, isolate the cavity.
%Except for the stainless steel external vacuum chamber, all the
%setup is made with aluminium and copper for high thermal
%conductivity to minimize temperature gradients.

\section{4.   Vibration Sensitivity Measurements}

For both cavities, the three vibration sensitivity components were
measured by shaking the cavity setup with sinusoidal signals in
the frequency range 1-10 Hz. Each cavity is housed in a vacuum
chamber supported on an optical table in two separate rooms. The
horizontal cavity setup (cavity, vacuum chamber and optical table)
is supported by an active vibration isolation platform. The cavity
itself is supported under vacuum with four 2 mm$^{2}$ Viton pads            %$^{\copyright}$
0.7 mm thick. The vertical cavity setup is isolated from vibration
using a passive isolation table, and is supported under vacuum
with the same Viton pads used for the horizontal setup. Air flow,
acoustic noise and large temperature fluctuations are strongly
filtered by containing the whole system in a thermo-acoustic
isolation box. The vacuum chamber temperature is actively
stabilized at $\approx 22^{\circ}$C.

\begin{figure}[b]
\includegraphics[width=0.5\textwidth]{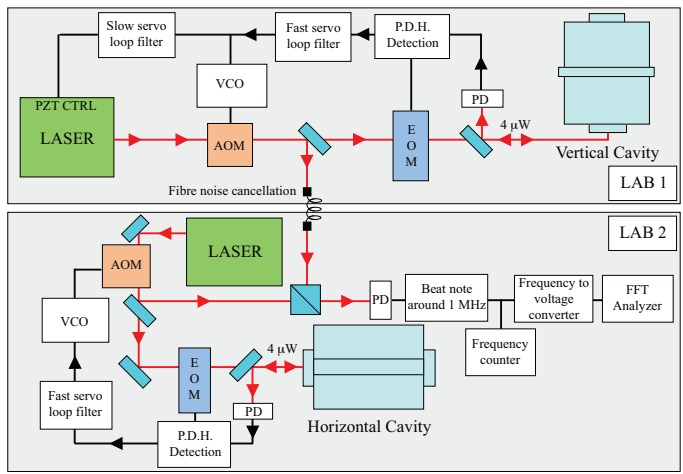}
\caption{ (Color online) Schematic showing the two independent
lasers used to create the beat-note with the horizontal and
vertical cavities. PDH: Pound Drever Hall \cite{pdh}, PD:
photodiode, FFT: fast Fourier transform, VCO: voltage controlled
oscillator, PZT CTRL: Piezo-electric transducer
control.\label{beat}}
\end{figure}

Two lasers in two different rooms (each with their own air
conditioning system) are independently stabilized to these two
cavities using the Pound-Drever-Hall technique. The beat-note
signal between the two stable lasers is demodulated by a
frequency-to-voltage converter and analyzed with a Fast Fourier
Transform analyzer (Fig. \ref{beat}).

\begin{figure}
\includegraphics[width=0.5\textwidth]{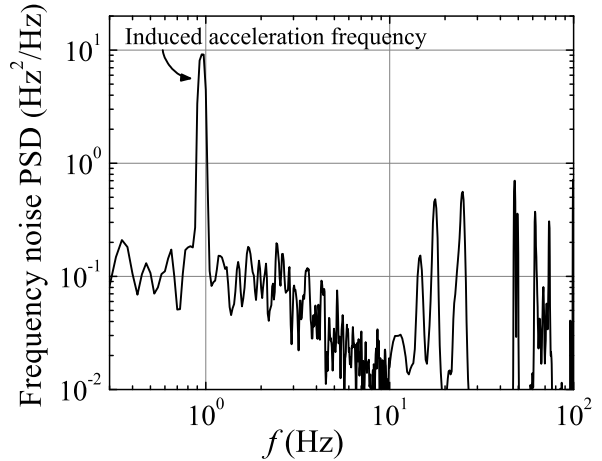}
\caption{Frequency noise power spectral density for induced
vertical acceleration in the horizontal cavity. Measurement
between two lasers locked separately onto the horizontal and the
vertical cavities. \label{vert}}
\end{figure}

A low noise seismometer placed on the top of the vacuum chamber is
used to measure the acceleration of the horizontal cavity in three
spatial directions. Each of the three orthogonal spatial
directions are excited in turn, while the amplitude of the induced
frequency tone and the strength of the acceleration is measured.
These measurements are iterated for several support-point
positions of the horizontal cavity. The active platform can apply
accelerations in a given direction with an amplitude of up to
$10^{-3}$m\,s$^{-2}$ rms. However, the coupling to other
directions could be as much as 10\%. A typical frequency response
measurement for induced vertical acceleration at 1 Hz can be seen
in Fig. \ref{vert}, with rms value is $\approx 7.5 \times 10
^{-4}$ m\,s$^{-2}$. The phase between the frequency response
signal (filtered and amplified) and the vibration excitation
signal is also measured, and gives the relative sign of the cavity
response. Measurement error bars have been estimated by
considering the contribution of the signal-to-noise ratio (5\%) of
the frequency deviation measurements, the acceleration crosstalk
effects (6\% to 10\% depending on the axis under consideration)
and the calibration error on vibration measurement (5\%).

These results for the horizontal cavity measuring the vertical
response agree with the vertically constrained model and contact
surfaces of about 1 mm$^2$ (Fig. \ref{vtcons}). The lowest
vibration sensitivity component observed is $5 \times 10^{-12}$
(m\,s$^{-2}$)$^{-1}$ and the dependence on the support-point
position is very low: $1.6 \times 10^{-12}$ (m\,s$^{-2}$)$^{-1}$
per mm.

The difference between the horizontal sensitivity component
measurements and simulations in Fig. \ref{horiz2} can be explained
by the unequal restoring forces of the four support-points on the
cavity. An asymmetry induces length variation and consequently the
coefficients $k_x^L$ and $k_y^L$ are no longer equal to zero. This
asymmetry is not reproducible when support point positions are
changed and can explain the dispersion of different measurements.
However, a linear fit of $k_x$ measured as function of $Y_p$ agree
with coefficients $k_x^T$ simulated for an offset between the
mechanical and optical axis of $d = 0.3$ mm. The longitudinal
vibration sensitivity is most sensitive to asymmetry of the
restoring forces. Components measured are four to eight times
larger than the coefficient $k_y^T$ simulated, even for a model
with a large offset of $d = 1$ mm (Fig. \ref{horiz2}).

A similar measurement has been performed to estimate the vibration
sensitivity component of the vertical cavity. In this case the
isolation platform is passive, so sinusoidal accelerations were
mechanically induced on the optical table. The results were
measured using a piezo-accelerometer, at a drive frequency of 1.2
Hz. All three axes were measured during the acceleration of each
direction to check for coupling. In this measurement, sensitivity
components obtained were $\sim 3.5 \times 10^{-12}$
(m\,s$^{-2}$)$^{-1}$ in the vertical and $1.4 \times 10^{-11}$
(m\,s$^{-2}$)$^{-1}$ in both horizontal directions. These results
are significantly better than those reported previously
\cite{ludlow2}.

Three other horizontal optical cavities were also constructed to
the design specifications presented in this paper. One cavity
operates at 698 nm and also has fused silica mirrors, showing
finesse of $\approx 600000$ and a fringe contrast better than
70\%. The other two are identical all ULE cavities operating at
1.55 $\mu$m.

\begin{figure}
\includegraphics[width=0.5\textwidth]{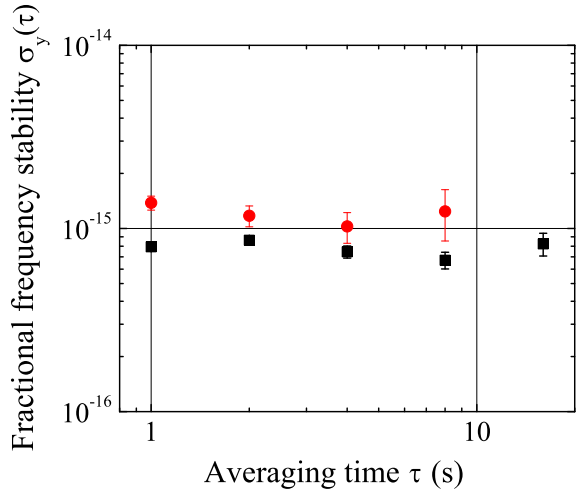}
\caption{  (Color online) Fractional frequency stability (Allan
standard deviation). Red circles plot: A laser locked onto a
vertical cavity at 1062.5 nm against the laser at 698 nm locked
onto another vertical cavity. Comparison realized via a
Ti:Sapphire-based frequency comb. Black squares plot (10
Hz\,s$^{-1}$ frequency drift removed): direct beat-note signal
between a laser locked onto the horizontal cavity and a laser
locked onto the vertical cavity, both at 1062.5 nm. \label{allan}}
\end{figure}

\section{5.   Frequency Stability Results}

\begin{figure}
\includegraphics[width=0.5\textwidth]{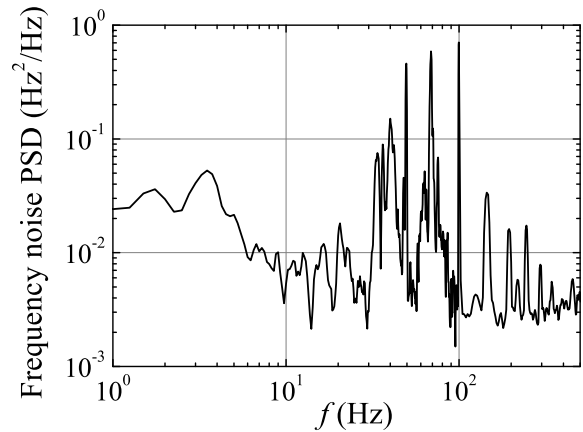}
\caption{Frequency noise power spectral density measured between
lasers locked separately onto the horizontal and the vertical
cavities. \label{freq}}
\end{figure}

A comparison between two independent lasers locked on the vertical
and the horizontal cavity has already shown the frequency
stability to be $7.9 \times 10^{-16}$@1s and $6.7 \times
10^{-16}$@8s (Fig. \ref{allan}, black squares plot). The frequency
drift is about 10 Hz\,s$^{-1}$ due to the incomplete thermal
control of the horizontal cavity, which will be largely reduced in
future by adding a thermal shield and actively controlling the
temperature for this set-up. Nevertheless, this demonstrates that
at least one of the two lasers exhibits a frequency stability
better than $5.6 \times 10^{-16}$@1s. A measurement of the
frequency noise of the beat-note (Fig. \ref{beat}) between lasers
stabilized onto the horizontal and vertical cavities is shown in
Fig. \ref{freq}.

These stability results for the horizontal and vertical cavities
can be compared with those from a different cavity built to the
same design, but with ULE mirrors rather than fused silica. This
laser is supported on 1.5 mm Viton pads, but otherwise has the
same set-up. Frequency stability for the all ULE cavity is $1.8
\times 10^{-15}$@1s \cite{jiang}, close to the expected thermal
noise limit. This demonstrates the improvement in stability
achieved by replacing ULE mirror substrate by fused silica mirror
substrate. The red circles plot in Fig. \ref{allan} is a
measurement of the relative frequency stability between a laser at
1062.5 nm locked onto the vertical fused silica mirror cavity, and
the 698 nm laser cavity locked onto the fused silica mirror
horizontal cavity. The comparison was realized via a Ti:Sapphire
based optical frequency comb \cite{ros}. The measured stability is
close to $10^{-15}$ from 1 to 10 s, a level at which we do not
exclude contributions from the Ti:Sapphire frequency comb noise.

\section{6.   Conclusion}

Two different optical cavity designs have been studied using
simulations with the purpose of decreasing the influence of
vibration on the length of the optical axis. These cavity designs
have been constructed and their vibration sensitivity measured. In
addition to the usual study of mirror translation, it is shown
that the effect of mirror tilt is of great significance.

For the horizontal cavity, vibration sensitivity is $\approx
10^{-11}$ (m\,s$^{-2}$)$^{-1}$ or better in all directions. The
vertical acceleration sensitivity component shows a small
dependence on the support-point positions of $1.6\times 10^{-12}$
(m\,s$^{-2}$)$^{-1}$ per mm. Therefore, fine tuning of their
positions is not necessary.
This is a very important improvement, since fine tuning is a time       %changed
consuming and delicate process. Frequency stability will be
improved further by optimizing the thermal environment of the
horizontal cavity to reduce the observed drift of 10 Hz\,s$^{-1}$.
Measurements also show that the vertical cavity has a low
vibration sensitivity, giving $3.5 \times 10^{-12}$
(m\,s$^{-2}$)$^{-1}$ in the vertical direction and $1.4 \times
10^{-11}$ (m\,s$^{-2}$)$^{-1}$ in the horizontal directions,
without any tuning.

The beat-note signal between two independent lasers stabilized to
these cavities (vertical and horizontal) shows a frequency
stability of $7.9 \times 10^{-16}$@1s and $6.7 \times 10^{-16}$@8s
(Fig. \ref{allan}, black squares plot). In contrast to previous
studies, the two systems are strongly independent (different
cavity design, different isolation systems, in different rooms)
ruling out the possibility of artificial improvements due to
correlation between the systems. Therefore, this result
demonstrates two ultra-stable lasers with stability in the
$10^{-16}$ range, lower than the noise floor of an all ULE cavity
with the same geometry, consequently this is the best result        %changed
achieved for a cavity of this length and compactness, potentially
suitable for applied systems. This is the first time such
stability has been measured in an ultra-stable cavity design
utilizing fused silica mirrors to reduce the thermal noise level.

\bibliographystyle{apsrev}
\bibliography{FinalCavities4}  %contents of FinalCavities.bbl can be pasted here to replace the bibliography

\end{document}